\documentclass[twocolumn,showpacs,preprintnumbers,amsmath,amssymb]{revtex4}

\usepackage{graphicx}
\usepackage{dcolumn}
\usepackage{bm}
\usepackage{epsfig}
\usepackage{graphics}
\usepackage{graphicx}
\usepackage{mathtext}

\begin{document}

\title{
Dynamic scaling, data-collapse and self-similarity in Barab\'{a}si-Albert networks
}%

\author{M. Kamrul Hassan$^{1}$, M. Zahedul Hassan$^{2}$, and Neeaj I. Pavel$^{1}$
}%
\date{\today}%

\affiliation{
$1$  Theoretical Physics Group, Department of Physics, University of Dhaka, Dhaka 1000, Bangladesh \\
$2$ Institute of Computer Science, Bangladesh Atomic Energy Commission, Dhaka 1000, Bangladesh 
}

\begin{abstract}%
In this article, we show that if each node of the Barab\'{a}si-Albert (BA) network is characterized by the generalized 
degree $q$, i.e. the product of their degree $k$ and the square root of their respective birth time, 
then the distribution function $F(q,t)$ exhibits dynamic scaling $F(q,t\rightarrow \infty)\sim t^{-1/2}\phi(q/t^{1/2})$ 
where $\phi(x)$ is the scaling function. We verified it by showing that a series of distinct $F(q,t)$ vs $q$ curves 
for different network sizes $N$ collapse onto a single universal curve if we plot $t^{1/2}F(q,t)$ vs $q/t^{1/2}$ instead. 
Finally, we show that the BA network falls into two universality classes depending on whether new nodes arrive with 
single edge ($m=1$) or with multiple edges ($m>1$).
\end{abstract}


\pacs{61.43.Hv, 64.60.Ht, 68.03.Fg, 89.75.Da}

\maketitle

Many complex systems can be described as an interwoven web of a large network if the constituents are regarded as nodes or 
vertices and the interactions between constituents as links or edges. 
For example, cells of living systems are networks of molecules linked by chemical interaction 
\cite{ref.brain,ref.protein,ref.metabolic}, 
the Internet is a network of routers and computers linked by cables or wireless connections \cite{ref.www}, the power-grid
 is a network of substations linked by transmission lines \cite{ref.powergrid}, 
and social networks are networks of individuals linked by friendships, professional ties, etc
\cite{ref.redner,ref.watts,ref.amaral,ref.newman,ref.newman_1}. The first theoretical attempt to guide
our understanding about complex network topology began with the seminal work of Paul Erd\"{o}s and Alfr\'{e}d R\'{e}nyi 
in 1959 \cite{ref.erdos}. The main result of the  Erd\"{o}s-R\'{e}nyi (ER) model 
is that the degree distribution $P(k)$, 
defined as the probability that a randomly chosen node is connected to $k$ other nodes by one edge,  
is Poissonian revealing that it is almost impossible to find nodes that have
significantly higher or fewer links than the average degree. However, real networks are neither 
completely regular where all the nodes have the same degree $k$ nor completely random where $P(k)$ is
Poissonian instead they are scale-free in character where $P(k)$ obeys power-law.
 
Just over a decade ago Barab\'{a}si and Albert revolutionized the notion of the network theory 
by recognizing the fact that natural and man-made networks are not static rather they grow 
by continuous addition of new nodes. They further argued that the new nodes 
establish links to the well-connected existing ones {\it preferentially} rather than {\it randomly} known as the 
preferential attachment (PA) rule. It essentially embodies the intuitive idea of the {\it rich get richer} 
principle of the Matthew effect in sociology \cite{ref.barabasi}. 
BA then presented a simple theoretical model 
incorporating both the ingredients and showed that the resulting 
network can reproduce the power-law degree distribution which most real life 
networks exhibit \cite{ref.barabasi_1,ref.review_1}. Recently, we have shown that random sequential 
partition of a square into contiguous and non-overlapping blocks can be described as a network with power-law 
degree distribution if blocks are regarded as nodes and common border between blocks as links \cite{ref.hassan}.       

A power-law distribution function is considered as scale-free since it looks the same 
regardless of the scale we look at it. In general, a function is called scale-free if it  
satisfies
\begin{equation}
\label{eq:1}
f(\lambda x)=g(\lambda)f(x).
\end{equation}
It can be rigorously proved that such function can only have none but power-law solution \cite{ref.newman_2}. 
The kinetic view of network topology and the new terminology, {\it scale-free
network}, has attracted physicists, mathematicians and computer scientists which resulted in a surge of research 
activities \cite{ref.barabasi_1,ref.review_1}.
On the other hand, a function $f(x,t)$ is said to obey dynamic scaling if one of the variable $t$ strictly denotes time 
and if it has the form
\begin{equation}
\label{eq:3}
f(x,t)\sim t^\theta \phi(x/t^z),
\end{equation}
where exponents $\theta$ and $z$ are fixed by the dimensional relations $[t^\theta]=[f]$ and $[t^z]=[x]$ respectively, 
while $\phi(\xi)$ is known as the scaling function \cite{ref.family_Vicsek}. 
There exists yet another scaling hypothesis, known as the finite-size scaling (FSS), that has been extensively 
used as a very 
powerful tool for estimating finite size effects in the critical phenomena. Within the FSS formalism, 
a function with exactly the same form as in Eq. (\ref{eq:3}) is said to obey finite-size scaling 
if $x$, though typically denoted by $\epsilon$, measures the distance from the critical point of the phenomenon 
under investigation and
$t$, though typically denoted by $L$, describes the linear size of the system \cite{ref.fss,ref.fss_1}. 

By definition, the BA model describes a time developing phenomenon and hence, besides its scale-free property, 
one could also look for its dynamic scaling property. This aspect of the BA model, however, has never been examined. 
To this end, we argue that each node in the dynamic network can be better characterized by generalized degree
$q$, the product of the square root of the birth time of each node and their corresponding degree $k$,
instead of the degree $k$ alone since the time of birth matters in the BA network. 
We find that the generalized degree distribution $F(q,t)$ has some non-trivial features and
exhibits dynamic scaling $F(q,t)\sim t^{-1/2}\phi(q/t^{1/2})$ in the long-time limit. We have
verified it using the idea of data-collapse.
For instance, we show that a series of distinct curves $F(q,t)$ versus $q$ for
different network sizes $N$ can be made to collapse onto a single universal curve
if we plot $t^{1/2}F$ against $q/t^{1/2}$. In addition, we find that it provides a means
to classify the BA networks into different universality classes which are otherwise regarded as the same. 
Establishing self-similarity and data-collapse or finding the scale-free property 
in any system has always been regarded as a significant progress towards gaining deeper insight into the problem 
\cite{ref.stanley,ref.barenblatt}.

The BA model begins with a small number of nodes $m_0$ 
as seeds which are already linked. Then, at each 
time step a new node with $m<m_0$ edges is added to the existing network. 
Edges of each new node are attached with $m$ different 
existing nodes by picking them preferentially with respect to their degree $k$. That is,
the probability that a new node will be connected to an already existing node $i$ is proportional to  
its degree $k_i$ and hence the degree $k_i$ of the node $i$ changes following the dynamical equation
\begin{equation}
\label{eq6}
{{\partial k_i}\over{\partial t}}=m{{k_i}\over{\sum_j^{N-1}k_j}}.
\end{equation}
Note that every time a node is added to the system it adds $m$
edges contributing to the increase of $2m$ degrees and hence  $\sum_j^{N-1}k_j=2m(t-1)$.  
Solving equation (\ref{eq6}) in the long time limit and using the fact that node $i$ is born at time $t_i$
with degree $m$, i.e. $k_i(t_i)=m$, gives
\begin{equation}
\label{eq8}
k_i(t)=m \Big ({{t}\over{t_i}}\Big )^\beta,
\end{equation}
where $\beta=1/2$. 
It implies that the degree of a node $i$ depends not only on the progressing time $t$, but also on the birth time
$t_i$. 

\begin{figure}
\includegraphics[width=5.5cm,height=8.5cm,clip=true,angle=-90]{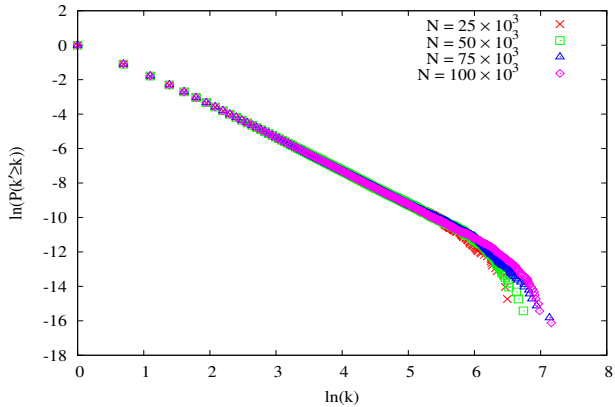}
\caption{Plots of the cumulative degree distributions $CD(k)\equiv P(k^\prime \geq k)$
is shown for the BA model with $m=1$. The data points are averaged over 500 independent realizations.
}
\label{fig 1}
\end{figure}%

We can find the degree distribution $P(k,t)$ by appreciating the fact that it is related to the 
homogeneous probability function $P(t_i)={{1}\over{t}}$ (nodes are added at equal time intervals) by
\begin{equation}
\label{eq9}
P(k)dk=-P(t_i)dt_i.
\end{equation}
Here, the minus sign is introduced to take into account the fact that the smaller the $t_i$ the larger the degree $k_i$ 
in the statistical sense. From equation (\ref{eq8-1}) we can easily find 
\begin{equation}
\label{eq8-1}
{{dt_i}\over{dk_i}}=-{{1}\over{\beta}} m^{1/\beta}tk_i^{-{{1+\beta}\over{\beta}}},
\end{equation}
and then substituting it as well as $P(t_i)={{1}\over{t}}$ into Eq. (\ref{eq9}) immediately gives
\begin{equation}
\label{eq10}
P(k)\sim 2m^2 k^{-\gamma} \hspace{0.5cm} {\rm with} \hspace{0.5cm} \gamma=3,
\end{equation}
since $\beta=1/2$. We thus find that $P(k)$ is independent of time $t$ albeit $k$ depends on $t$ and it satisfies the 
scale-free form given by Eq. (\ref{eq:1}). 
It means that the network in the long time limit self-organizes into a scale-free state
where it no longer depends on time.

It is interesting to note that the degree distribution $P(k)$ typically has a
long tail with relatively scarce data points which turns into a fat-tail 
when we plot $P(k)$ in log-log scale. This complicates the process of identifying the range over which the
power-law holds and hence estimating the exponent $\gamma$. One way of reducing the 
noise at the tail-end is to plot   
cumulative distribution $P(k^\prime \geq k)$ which is related to degree distribution $P(k)$ via
\begin{equation}
P(k)=-{{dP(k^\prime\geq k)}\over{dk}}.
\end{equation}
Figure 1 shows that $\ln (P(k^\prime \geq k))$ decays linearly against $\ln (k)$ 
with slope equals to $\gamma-1=2$ as expected according to equation (\ref{eq10})
except at the tail-end due to finite-size effect. 
Furthermore, it shows that the extent of linearity increases as network size $N$ increases
revealing that the sudden fall off near the tail-end is indeed due to finite-size effect. 
Waclaw and  Sokolov in \cite{ref.waclaw} suggested
that the finite size effect in the degree distribution of the BA network can be well approximated by $P_N(k)=P(k)w(k/\sqrt{N})$ where
the cut-off function $w(x)$ is found highly sensitive to both $m$ and $m_0$.

\begin{figure}
\includegraphics[width=5.5cm,height=8.5cm,clip=true,angle=-90]{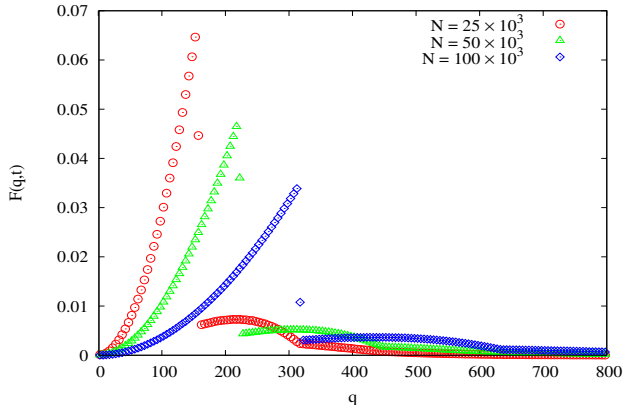}
\caption{Generalized degree distributions $F(q,t)$ for $m=1$ are shown against $q$ for three different network size $N$.
In each case data in the graph represent averaged over $500$ independent realizations.
}
\label{fig2}
\end{figure}%

We thus find that the degree distribution $P(k,t)$ self-organizes into a time invariant
state in the long-time and large-size limit instead of a state where it assumes a dynamic scaling form.
We therefore have to look for another quantity which exhibits dynamic scaling in the BA model.
Note that a complete characterization of the nodes in the BA network can be made by specifying
their degree $k$ as well as their respective birth time. We, however, 
find it highly instructive to combine the two into a single variable using Eq. (\ref{eq8}). That is, 
the node $i$ at time $t$ can be characterized by the generalized degree $q_i(t)= k_it_i^\beta$ 
where the exponent $\beta$ value is fixed by the dimensional relation $[q^{1/\beta}]=[t]$.
The advantage of using $q$ is that it depends only on time $t$ since according to Eq. (\ref{eq8}) we have 
\begin{equation}
\label{eq11}
q_i(t)\sim t^\beta  \hspace{0.5cm} {\rm with} \hspace{0.5cm} \beta=1/2. 
\end{equation}
It is now customary to consider the generalized degree distribution 
$F(q,t)$ instead of the traditional degree distribution $P(k)$.  
That is, if we pick a node at random at time $t$ then $F(q,t)$ is the probability that its generalized 
degree is $q$. In Fig. 2 we have drawn $F(q,t)$ vs $q$ 
for three different network sizes $N$ and found some remarkable features. For instance, 
we find that the value of $F(q,t)$ 
initially increases quite sharply and then register a sudden and sharp fall to a non-zero value at $q_c$ from which it 
rises again to a secondary maximum followed by a smooth
decrease with a long tail. The value $q_c$, where the first minimum occurs, 
increases with the network size $N$ and at the same time
both primary and secondary heights systematically decreases as $N$ increases.

We shall now invoke the idea of the dimensional analysis to show that $F(q,t)$ in the long-time large-size limit 
self-organizes into a state where it exhibits dynamic scaling \cite{ref.barenblatt}.
Clearly, there are two governing parameters $q$ and $t$ and a governed parameter $F(q,t)$ in the problem at hand. 
However, according to 
Eq. (\ref{eq11}) the governing parameter $q$ can be expressed in terms of time $t$ alone 
and hence we can define a dimensionless governing parameter 
\begin{equation}
\label{eq:12}
\xi={{q}\over{t^{1/2}}}.
\end{equation} 
It means we can express $F(q,t)$ too in terms of time $t$ alone since time $t$ 
is chosen to be an independent parameter. Applying the
power-monomial law of the dimensional function of a physical quantity we can write a dimensional relation 
$F(q,t)\sim t^{\alpha}$ where the exponent $\alpha$ assumes a value that makes
$t^{\alpha}$ bear the dimension of $F(q,t)$ \cite{ref.barenblatt}. We can therefore define 
yet another dimensionless governed parameter $\phi$ as follows
\begin{equation}
\label{eq:13}
\phi=\frac{F(q,t)}{t^{\alpha}}.
\end{equation} 
Now, within a given class one can pass from one unit of measurements to another 
by changing $t$, for instance, by an arbitrary factor leaving the other factor $q$ unchanged. Upon such transition 
within the given class the 
numerical value of the quantity on the right hand side of Eq. (\ref{eq:13}) must remain unchanged
since the left hand side is a dimensionless quantity. 

\begin{figure}
\includegraphics[width=5.5cm,height=8.5cm,clip=true,angle=-90]{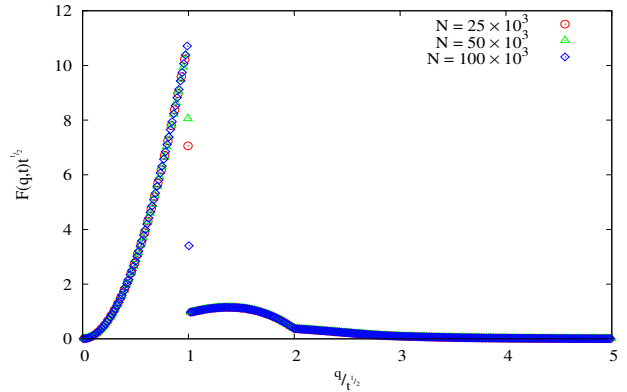}
\caption{The same data of Fig. 2 for $m=1$ is shown in the self-similar coordinates $t^{1/2}F(q,t)$ and $q/t^{1/2}$ and
we find that all the three curves of Fig 2 collapsed onto a single universal curve.
}
\label{fig 3}
\end{figure}%

\begin{figure}
\includegraphics[width=5.5cm,height=8.5cm,clip=true,angle=-90]{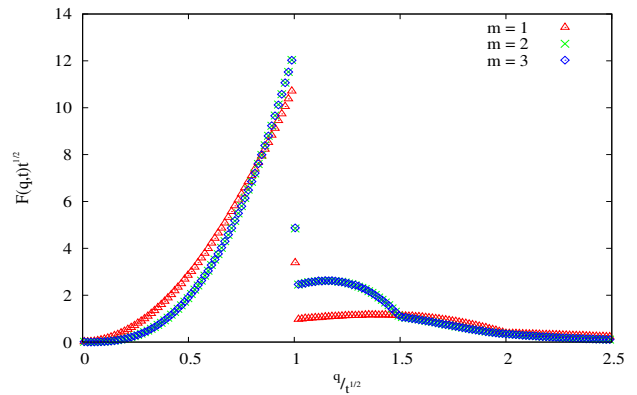}
\caption{The value of $F(q,t)$ for $m=1,2$ and $3$ are plotted in the self-similar coordinates. It clearly shows that
universal curve for $m=1$ do not collapse with those for $m>1$ and hence belong to two different classes.
}
\label{fig 4}
\end{figure}

We thus find that the quantity $\frac{F(q,t)}{t^{\alpha}}$ {\it vis-a-vis} $\phi$ can at best be a function of
another dimensionless quantity $\xi$ given by Eq. (\ref{eq:12}) since this is the only dimensionless governing parameter.
We can express Eq. (\ref{eq:13}) as 
\begin{equation}
\label{eq:ansatz}
F(q,t)\sim t^{\alpha}\phi(q/t^{1/2}),
\end{equation}
where the exponent $\alpha$ is obtained by applying the normalization condition $\int_0^\infty F(q,t)dq=1$ to give $\alpha=-1/2$. We thus finally find that the generalized degree distribution
assumes exactly the same dynamical scaling form as Eq. (\ref{eq:3}). 
An interesting aspect of the structure of the dynamic scaling form given by Eq. (\ref{eq:ansatz}) is that the 
distribution function $F(q,t)$ at various moments of time can be obtained 
from one another by a similarity transformation
\begin{equation}
q\longrightarrow \lambda^{1/2} q, \hspace{0.20cm} t\longrightarrow \lambda t, \hspace{0.20cm} F\longrightarrow \lambda^{-1/2} F,
\end{equation}
revealing the self-similar nature of the function $F(q,t)$.

The question is: How do we verify Eq. (\ref{eq:ansatz}) using the data extracted from numerical simulation? 
The best way of verifying it is to invoke the idea of data-collapse. Note that according to Eqs. (\ref{eq:12}) and
(\ref{eq:13}) the quantities
$t^{1/2}F(q,t) and \phi(q/t^{1/2})$ are both dimensionless and hence the data of $F(q,t)$ for various 
network size $N$ should collapse on a single universal curve 
if we plot $t^{1/2}F(q,t)$ against $q/t^{1/2}$ which are known 
as the self-similar coordinates. 
In Fig. 3 we have drawn the scaled generalized degree distribution $t^{1/2} F(q,t)$ as a function of scaled 
generalized degree $q/t^{1/2}$ for three different network sizes 
$N$ considering $m=1$ in each case and find that all the distinct curves of Fig. 2 merge superbly onto a single curve
as expected. It clearly well confirms the validity of the dynamic scaling.
The first discontinuity in Fig. 3 occurs at $q/t^{1/2}=1$ and the second discontinuity at $q/t^{1/2}=2$ and both
are quite sharp. A careful observation also reveals that there exists a third discontinuity at $q/t^{1/2}=3$ which
seems quite weak and hard to notice.  
Now the question is: What if nodes arrive in the BA network with more than one edges ($m>1$)?
Below, we attempt to give an answer to this question.

\begin{figure}
\includegraphics[width=5.5cm,height=8.5cm,clip=true,angle=-90]{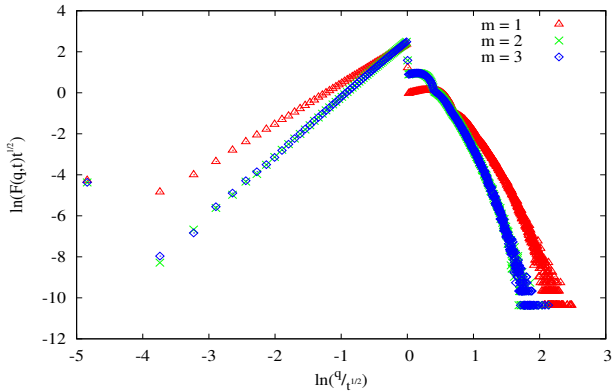}
\caption{Plots of $\ln (t^{1/2}F(q,t)$ vs $\ln (q/t^{1/2})$ for $m=1$ and $m>1$. It shows
that the scaling functions $\phi(\xi)$ in both cases grow obeying power-law until $q/t^{1/2}<1$ but with exponent $2$ and
$2.9$ for $m=1$ and $m>1$ respectively.
}
\label{fig 5}
\end{figure}%

\begin{figure}
\includegraphics[width=5.5cm,height=8.5cm,clip=true,angle=-90]{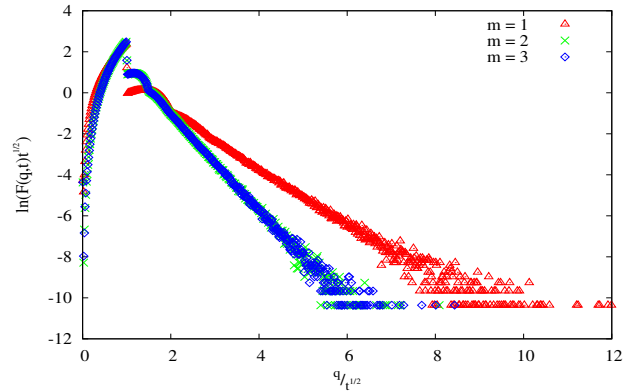}
\caption{Plots of $\ln (t^{1/2}F(q,t)$ as a function of $\xi=q/t^{1/2}$ for $m=1$ and $m>1$ which clearly show that
the scaling function $\phi(\xi)$ beyond $q/t^{1/2}=2$ and $q/t^{1/2}=1.5$ respectively decay exponentially
but with two different decay constant.}
\label{fig 6}
\end{figure}%

According to our dimensional analysis it is expected that the data points for 
the BA network that grow by sequential addition of a node with multiple edges ($m>1$) should also lie on the same curve 
as the ones for single edge ($m=1$) unless they are fundamentally different in some subtle way which the ordinary degree 
distribution $P(k)$ fail to differentiate. Surprisingly, we find that the data points of 
$t^{1/2}F(q,t)$ for $m=2, 3, 4$ etc. do collapse on a single curve if plotted against $q/t^{1/2}$ but do not coincide
with the one for $m=1$ (see Fig. 4). Although the two distinct universal curves, one for $m=1$ and the other for $m>1$, 
behave qualitatively almost in the similar fashion they are however quantitatively different. For
instance, one noteworthy difference is that although the first discontinuity
occurs at the same point as for $m=1$, the second one, however, occurs at $q/t^{1/2}=1.5$ for $m>1$ while 
at $q/t^{1/2}=2$ for $m=1$. These differences can be even better appreciated 
if  we plot $t^{1/2}F(q,t)$ vs $q/t^{1/2}$ either in the log-log scale or in the log-linear 
scale as shown in figures 5 and 6 
respectively. Using these figures, we can write the solutions for the universal scaling function $\phi(\xi)$ both for $m=1$ 
 \begin{equation}
\label{eq:binary_2}
\phi(\xi) \sim \left \{ \begin{array}{r@{\qquad\qquad}l}                  
\xi^2 \hfill  &   \hspace{0.5cm}{\rm if} \hspace{.4cm} \xi<1 \\
e^{-a\xi} \hfill & \hspace{0.5cm} {\rm if} \hspace{.4cm} \xi>2 \\
\end{array} \right., 
\end{equation}
and for $m>1$ 
\begin{equation}
\label{eq:binary_3}
\phi(\xi) \sim \left \{ \begin{array}{r@{\qquad\qquad}l}                  
\xi^{2.9} \hfill  &   \hspace{0.5cm}{\rm if} \hspace{.4cm} \xi<1 \\
e^{-b\xi} \hfill & \hspace{0.5cm} {\rm if} \hspace{.4cm} \xi>1.5 \\
\end{array} \right., 
\end{equation}
where $a \approx 1.4$ and $b\approx 2.5$.

Beside proving the existence of self-similarity by data-collapse, we have also found that 
there are two distinct classes of networks resulting from the BA model. 
(i) The network that grows by sequential
addition of one node with single edge ($m=1$) and (ii) the network that grows by sequential addition of one node
with multiple edges ($m>1$). There must be some reasons behind such behaviour. 
One apparent difference between the two classes of networks is that the clustering coefficient $C_N=0$ if $m=1$ and 
$C_N\neq 0$ rather decays like 
$C_N\sim N^{-0.75}$ if $m>1$ for network of size $N$. 
Regardless of the value $m$, the universal scaling function $\phi(\xi)$ always suffers two discontinuities
which divide the nodes in the network always into three different classes. For instance, 
nodes which have generalized degree $q<t^{1/2}$ fall into one class,
those which have generalized degree within $t^{1/2}<q<2t^{1/2}$ for $m=1$ and $t^{1/2}<q<1.5t^{1/2}$ for $m>1$
into another class and finally those which lie beyond these limits.

To summarize, we have proposed a systematic processing procedure to obtain the dynamic scaling ansatz and the self-similar coordinates
for the BA networks. To this end we have used dimensional analysis and shown that it provides deeper insight into the problem. 
We have found that if the nodes of the BA network are characterized by the generalized degree $q_i(t)$ of the
node $i$ at time $t$, defined as the product of its degree $k_i(t)$ at time $t$ and the square root of its birth time $t_i$,
then its distribution $F(q,t)$ in the long-time limit exhibits dynamic scaling $F(q,t)\sim t^{-1/2}\phi(q/t^{1/2})$.  
We have verified it numerically by showing that all the data points for various network size $N$ collapse onto a single
universal curve if we plot $t^{1/2}F(q,t)$ as a function of $q/t^{1/2}$. Our findings suggest that there are two distinct 
classes of universal scaling functions, depending on whether each new node arrive with single edge or 
with multiple edges. The scaling functions $\phi(\xi)$ for $m=1$ and $m>1$ 
have some remarkable features in the sense that they both suffer discontinuity once at $\xi=1$ and 
then the curve for $m=1$ at $\xi=2$ and the corresponding curve for $m>1$ at $\xi=1.5$ 
albeit qualitatively they are very much similar.
The two types of BA networks are indeed fundamentally different. For instance, the clustering coefficient $C_N=0$ 
if $m=1$ and obeys the same power-law $C_N\sim N^{-0.75}$ if $m>1$. The idea of data-collapse developed for the BA model could also be applied 
on real life data as well as on the existing dynamic networks.  
It would certainly be an attractive proposition to check if the generalized degree distribution
derived from the real life data or from other kinetic network too exhibit 
dynamic scaling with similar universal features or not. We intend to continue our work
in this direction in our future endeavour. 

MKH and NIP gratefully acknowledge financial support from the Bose Centre for Advanced Study and Research of Dhaka
University, Bangladesh.


\end{document}